\documentclass[12pt]{article}
\title{\bf Aharonov-Bohm Effect in the\\
Abelian-Projected $SU(3)$-QCD with $\Theta$-term}
\author{Dmitri Antonov \thanks{Permanent address:
Institute of Theoretical and Experimental Physics, 
B. Cheremushkinskaya 25, RU-117 218 Moscow, Russia.}{\,}
\thanks{E-mail address: {\tt antonov@mailbox.difi.unipi.it}} 
\\
{\it INFN-Sezione di Pisa, Universit\'a degli studi di Pisa,}\\
{\it Dipartimento di Fisica, Via Buonarroti, 2 - Ed. B - 56127 Pisa, Italy}}
\date{}
\begin{document}
\maketitle
\vspace{1mm}
\centerline{\bf {Abstract}}
\vspace{3mm}
\noindent
By making use of the path-integral duality transformation,
string representation of the Abelian-projected $SU(3)$-QCD with 
the $\Theta$-term is derived. 
Besides the short-range (self-)interactions of quarks (which due to the 
$\Theta$-term acquire a nonvanishing 
magnetic charge, 
{\it i.e.} become dyons) and electric Abrikosov-Nielsen-Olesen 
strings, the resulting effective action contains also 
a long-range topological 
interaction of dyons with strings. 
This interaction, which has the form of the 4D Gauss linking number 
of the trajectory of a dyon with the world-sheet of a closed string, 
is shown to become nontrivial at 
$\Theta\ne 3\pi\times{\,}{\rm integer}$. At these values of $\Theta$, 
closed electric Abrikosov-Nielsen-Olesen strings in the model under study 
can be viewed as solenoids scattering dyons, which is the 4D analogue 
of the Aharonov-Bohm effect. 

\vspace{3mm}

\vspace{3mm}

\newpage

\section{Introduction}

Recently, by making use of the method of Abelian projections~\cite{th}, 
a remarkable progress in the 
description of confinement in QCD has been achieved (for a recent review 
see~\cite{digiacomo}). In this way, most of the 
results have been obtained for the simplest case of the $SU(2)$-group.
That is because in this case 
the respective Cartan subgroup is just $U(1)$. Owing to this fact,    
under the assumptions on the irrelevance of the off-diagonal 
degrees of freedom in the IR limit (the so-called Abelian dominance 
hypothesis)~\cite{abdom} and condensation of monopoles arising 
during the Abelian projection, the resulting effective 
theory takes the form of the dual Abelian Higgs model.
In the latter model, confinement of external electrically charged 
particles is provided by the formation of electric Abrikosov-Nielsel-Olesen
(ANO) strings~\cite{ano} between these particles. This is just the 
essence of the 't Hooft-Mandelstam scenario of confinement~\cite{mand}.

Besides this simplest $SU(2)$-case, an effective $[U(1)]^2$ gauge invariant 
Abelian-projected 
theory of the realistic $SU(3)$-QCD, based on the Abelian dominance 
hypothesis, has also been proposed~\cite{maedan}. 
Quite recently, confining properties of this model were investigated 
by a derivation of its string representation~\cite{su3} (for a review 
see~\cite{rev}). Moreover, in Ref.~\cite{mpla} the collective effects in the 
grand canonical ensemble of strings in this model were investigated. 
In particular, a crucial difference of this ensemble from 
the one of the $SU(2)$ Abelian-projected theory~\cite{ijmpa} 
(following from the appearance of two types of (self-)interacting strings 
in the $SU(3)$-case) has been found. 

The present Letter is
devoted to the investigation of Abelian-projected $SU(3)$-QCD 
extended by the introduction of the $\Theta$-term. 
The relevance of this term to the instanton physics is a well 
known issue, and we will not discuss it further, referring the 
reader to classical monographs (see {\it e.g.}~\cite{mono}). 
Within our investigations, it will be 
demonstrated that the introduction of the $\Theta$-term leads 
to the appearance of a long-range topological interaction of 
quarks (which due to this term become dyons by 
acquiring a nonvanishing magnetic charge~\cite{witten}) with 
the world-sheets of closed electric ANO strings. This interaction  
has the form of the Gauss linking number of these objects 
and can be interpreted as the 4D analogue~\cite{four} of the Aharonov-Bohm 
effect~\cite{ab}, where strings play the r\^ole of solenoids,
which scatter dyons. In particular, it will be demonstrated 
that at certain discrete critical values of $\Theta$ this effect 
disappears. This can be interpreted in such a way 
that at these values of $\Theta$, the 
relation between the electric flux in the ANO string (viewed as a 
solenoid) and the magnetic charge of the dyon is so, that the 
scattering of dyons over strings is absent. 
Note that the 
long-range interaction of dyons with magnetic ANO strings 
in the Abelian Higgs model (which corresponds to the theory 
dual to the Abelian-projected $SU(2)$-QCD with the $\Theta$-term)
has been studied in Ref.~\cite{emil}. Similar interaction of 
an external electrically charged particle with magnetic ANO strings 
in the same model but without $\Theta$-term has 
been found in Ref.~\cite{polikarp} by the evaluation 
of the Wilson loop describing this particle.

It will be also shown that, as it can be expected from the very beginning,  
besides this long-range interaction of strings and dyons, 
the resulting effective action contains the short-range 
(self-)interactions 
of these objects. The short-rangeness of all these 
interactions is due to the (self-)exchanges of strings and dyons 
by massive dual gauge bosons.
Among these short-range interactions, the 
selfinteraction of the open world-sheets of strings, which end up at 
dyons, is the most important one, since it is this interaction, which 
provides the linear growth of the confining dyon-antidyon potential.

The organization of the Letter is as follows. 
In the next Section, we shall discuss an effective 
theory of the Abelian-projected $SU(3)$-QCD with the $\Theta$-term.
In Section 3,
a derivation of the announced 
string representation for the partition function of this model
will be presented, and the obtained results will be briefly discussed.
Finally, in three Appendices, some technical details of calculations, 
which are necessary for a derivation of the formulae 
from the main text, are outlined.

\section{The Model}

In the present Section, we shall discuss the model, whose 
string representation will be derived below.
To start with, let us consider the partition function of an 
effective $[U(1)]^2$ gauge invariant Abelian-projected theory 
of the pure $SU(3)$-gluodynamics ({\it i.e.} a theory without quarks and 
the $\Theta$-term)~\cite{maedan}.
It reads~\footnote{
Throughout the present Letter, 
all the investigations will be performed in the Euclidean space-time.}

$$
{\cal Z}=\int \left|\Phi_a\right| {\cal D}\left|\Phi_a\right|
{\cal D}\theta_a {\cal D}{\bf B}_\mu
\delta\left(\sum\limits_{a=1}^{3}
\theta_a\right)\times$$

\begin{equation}
\label{et6}
\times\exp\left\{-\int d^4x\left[\frac14{\bf F}_{\mu\nu}^2+
\sum\limits_{a=1}^{3}\left[\left|\left(\partial_\mu-
ig_m{\bf e}_a{\bf B}_\mu\right)\Phi_a\right|^2+
\lambda\left(|\Phi_a|^2-\eta^2\right)^2\right]\right]\right\}.
\end{equation}
Here, $\Phi_a=\left|\Phi_a\right|{\rm e}^{i\theta_a}$ are the 
dual Higgs fields, which describe the condensates of monopoles 
emerging after the Abelian projection, and $g_m$ is the magnetic 
coupling constant related to the 
electric one via the topological 
quantization condition $g_mg=4\pi k$. In what follows,
we shall for simplicity restrict ourselves to the monopoles
possessing the minimal charge only, {\it i.e.} set $k=1$.
Next, ${\bf B}_\mu$ is the ``magnetic'' potential dual 
to the ``electric'' potential 
${\bf a}_\mu\equiv\left(A_\mu^3, A_\mu^8\right)$, 
where $A_\mu^{3,8}$ are just the diagonal components of the gluonic 
field. 
On the R.H.S. of Eq.~(\ref{et6}), there have also been introduced 
the so-called root vectors ${\bf e}_a$'s, which have the form 

$$
{\bf e}_1=\left(1,0\right),~ 
{\bf e}_2=\left(-\frac12,-\frac{\sqrt{3}}{2}\right),~ 
{\bf e}_3=\left(-\frac12,\frac{\sqrt{3}}{2}\right).$$ 
These vectors play the r\^ole of the structural constants 
in the first of the following commutation relations 

$$\left[{\bf H},E_{\pm a}\right]=\pm{\bf e}_aE_{\pm a},~ 
\left[E_{\pm a},E_{\pm b}\right]=\mp\frac{1}{\sqrt{2}}
\varepsilon_{abc}E_{\mp c},~ 
\left[E_a,E_{-b}\right]=\delta_{ab}{\bf e}_a{\bf H}.$$
In these relations,  
${\bf H}\equiv\left(H_1,H_2\right)=\left(T^3,T^8\right)$ are 
the diagonal $SU(3)$-generators, which generate the Cartan 
subalgebra, where from now on $T^i=\frac{\lambda^i}{2}$, $i=1,\ldots,8$,   
are just the $SU(3)$-generators with $\lambda^i$'s denoting 
the Gell-Mann matrices. 
We have also 
introduced the so-called step operators $E_{\pm a}$'s (else called 
raising operators for positive $a$'s and lowering operators otherwise) 
by redefining the rest (non-diagonal) $SU(3)$-generators as follows 

$$E_{\pm 1}=\frac{1}{\sqrt{2}}\left(T^1\pm iT^2\right),~ 
E_{\pm 2}=\frac{1}{\sqrt{2}}\left(T^4\mp iT^5\right),~ 
E_{\pm 3}=\frac{1}{\sqrt{2}}\left(T^6\pm iT^7\right).$$
(Clearly, these operators are non-Hermitean in the sense that 
$(E_a)^\dag=E_{-a}$.) It is also worth noting that owing to the fact that 
the original $SU(3)$ group is special, the phases $\theta_a$'s of the 
dual Higgs fields are related to each other by the constraint 
$\sum\limits_{a=1}^{3}\theta_a=0$. This constraint was imposed by introducing 
the corresponding $\delta$-function into the R.H.S. of Eq.~(\ref{et6}).
 
In what follows, we shall be interested in the study of the 
model~(\ref{et6}) in the London limit, {\it i.e.} the limit 
of infinitely large coupling constant $\lambda$ of the dual Higgs fields  
(Clearly, this limit corresponds to the case of 
infinitely large masses of the dual Higgs bosons.). It is just this limit, 
where the model under study allows for an exact reformulation 
in terms of the integral over closed electric 
ANO strings~\cite{su3}~\footnote{Note that the size 
of the core of the string (vortex in 3D) is equal to the 
inverse mass of the dual Higgs fields, which means 
that the London limit corresponds 
to infinitely thin strings.}. In the London limit, the radial 
parts of the Higgs fields can be integrated out, and the 
partition function~(\ref{et6}) takes the form 

$$
{\cal Z}=\int {\cal D}{\bf B}_\mu {\cal D}\theta_a^{\rm sing.}
{\cal D}\theta_a^{\rm reg.} {\cal D}k\delta\left(\sum\limits_{a=1}^{3}
\theta_a^{\rm sing.}\right)\times
$$

\begin{equation}
\label{suz2}
\times\exp\Biggl\{\int d^4x\Biggl[
-\frac14{\bf F}_{\mu\nu}^2-\eta^2\sum\limits_{a=1}^{3}
\left(\partial_\mu\theta_a-g_m{\bf e}_a{\bf B}_\mu\right)^2+
ik\sum\limits_{a=1}^{3}\theta_a^{\rm reg.}\Biggr]\Biggr\}. 
\end{equation}
Since in the model~(\ref{et6}) there exist   
string-like singularities of the 
ANO type, in Eq.~(\ref{suz2}) 
we have decomposed the total phases 
of the dual Higgs 
fields into multivalued and singlevalued (else oftenly called singular and 
regular, respectively) parts, $\theta_a=
\theta_a^{\rm sing.}+
\theta_a^{\rm reg.}$, and imposed the constraint of vanishing of the 
sum of regular parts by introducing the integration over the 
Lagrange multiplier $k(x)$. Analogously to the 
dual Abelian Higgs model, in the 
model~(\ref{suz2}), 
$\theta_a^{\rm sing.}$'s 
describe a certain electric string configuration and 
are related to the closed world-sheets 
$\Sigma_a$'s of strings of three types 
via the equation 

\begin{equation}
\label{suz3}
\varepsilon_{\mu\nu\lambda\rho}\partial_\lambda\partial_\rho
\theta_a^{\rm sing.}(x)=2\pi\Sigma_{\mu\nu}^a(x)\equiv
2\pi\int\limits_{\Sigma_a}^{}d\sigma_{\mu\nu}\left(x^{(a)}(\xi)\right)
\delta\left(x-x^{(a)}(\xi)\right).
\end{equation}
This equation is nothing else, but the covariant formulation 
of the 4D analogue of the Stokes theorem for the gradient 
of the field $\theta_a$, written in the local form. In Eq.~(\ref{suz3}), 
$x^{(a)}(\xi)\equiv x_\mu^{(a)}(\xi)$ is a vector, which parametrizes 
the world-sheet $\Sigma_a$ with $\xi=(\xi^1, \xi^2)$
standing for the 2D coordinate. On the other hand, 
the regular parts of the phases, $\theta_a^{\rm reg.}$'s, 
describe a singlevalued fluctuation around the above mentioned 
given string configuration. 
Note that owing to the one-to-one correspondence between 
$\theta_a^{\rm sing.}$'s and $\Sigma_a$'s, established by 
Eq.~(\ref{suz3}),
the integration over $\theta_a^{\rm sing.}$'s is implied 
in the sense of a certain prescription of the summation   
over string world-sheets. One of the possible 
concrete forms of such a prescription, corresponding to the summation 
over the grand canonical ensemble of 
virtual pairs of strings with opposite winding numbers, has been 
considered in Refs.~\cite{mpla, ijmpa}.
It is also worth noting that due to Eq.~(\ref{suz3}) 
the integration measure over the full phases 
$\theta_a$'s can be shown to become 
factorized into the product of measures over the 
fields $\theta_a^{\rm sing.}$'s and 
$\theta_a^{\rm reg.}$'s.

As it was announced in the Introduction, the model 
whose string representation will be of our interest
is not the London limit of 
pure Abelian-projected $SU(3)$-gluodynamics~(\ref{suz2}), 
but rather the full $SU(3)$-QCD extended by the $\Theta$-term.
For a certain quark colour $c=R, B, G$ (red, blue, green, 
respectively), the partition function of such a theory reads~\cite{maedan} 

$$
{\cal Z}_c=\int {\cal D}{\bf B}_\mu {\cal D}\theta_a^{\rm sing.}
{\cal D}\theta_a^{\rm reg.} {\cal D}k\delta\left(\sum\limits_{a=1}^{3}
\theta_a^{\rm sing.}\right)
\exp\Biggl\{\int d^4x\Biggl[
-\frac14\left({\bf F}_{\mu\nu}+{\bf F}_{\mu\nu}^{(c)}\right)^2-
$$

\begin{equation}
\label{partf}
-\eta^2\sum\limits_{a=1}^{3}
\left(\partial_\mu\theta_a-g_m{\bf e}_a{\bf B}_\mu\right)^2+
ik\sum\limits_{a=1}^{3}\theta_a^{\rm reg.}
+\frac{i\Theta g_m^2}{16\pi^2}
\left({\bf F}_{\mu\nu}+{\bf F}_{\mu\nu}^{(c)}\right)
\left(\tilde{\bf F}_{\mu\nu}+\tilde{\bf F}_{\mu\nu}^{(c)}\right)
\Biggr]\Biggr\}. 
\end{equation}
Here, $\tilde {\cal O}_{\mu\nu}\equiv\frac12\varepsilon_{\mu\nu\lambda\rho}
{\cal O}_{\lambda\rho}$, and 
${\bf F}_{\mu\nu}^{(c)}$ is the field strength tensor 
of a test quark of the colour $c$, which moves along a certain 
contour $C$. This field strength tensor obeys the equation 
$\partial_\mu\tilde {\bf F}_{\mu\nu}^{(c)}={\bf Q}^{(c)}j_\nu$,
where $j_\mu(x)=\oint\limits_{C}^{}dx_\mu(\tau)\delta(x-x(\tau))$, 
and the vectors of colour charges read 

$${\bf Q}^{(R)}=\left(\frac{g}{2}, \frac{g}{2\sqrt{3}}\right),~ 
{\bf Q}^{(B)}=\left(-\frac{g}{2}, \frac{g}{2\sqrt{3}}\right),~ 
{\bf Q}^{(G)}=\left(0, -\frac{g}{\sqrt{3}}\right).
$$
Clearly, ${\bf F}_{\mu\nu}^{(c)}={\bf Q}^{(c)}\tilde{\cal F}_{\mu\nu}$, 
where either 

\begin{equation}
\label{calF}
{\cal F}_{\mu\nu}(x)=-\frac{1}{4\pi^2}\left[\frac{\partial}{\partial x_\mu}
\int d^4y\frac{j_\nu(y)}{(x-y)^2}-
\frac{\partial}{\partial x_\nu}
\int d^4y\frac{j_\mu(y)}{(x-y)^2}\right]
\end{equation}
or ${\cal F}_{\mu\nu}=-\Sigma_{\mu\nu}$.
Here, $\Sigma_{\mu\nu}(x)=\int\limits_{\Sigma}^{}d\sigma_{\mu\nu}
(x(\xi))\delta(x-x(\xi))$ is the vorticity tensor current associated 
with the world-sheet $\Sigma$ of 
a certain open electric ANO string, bounded by the contour $C$.

\section{Aharonov-Bohm Effect and Confinement}

In order to proceed with the string representation of the 
model~(\ref{partf}), it is useful to employ 
the so-called path-integral duality transformation 
(see {\it e.g.} Ref.~\cite{lee}, where this transformation 
is presented for the simplest case of the (dual) Abelian Higgs model),
whose details are outlined in Appendix A. The result 
of this procedure reads~\footnote{From now on, we omit everywhere 
the normalization constant, implying for every colour $c$ 
the normalization condition
${\cal Z}_c[C=0]=1$.} 

$$
{\cal Z}_c=\exp\left[-\frac{(\Theta g_m)^2}{12\pi^2}\int d^4x
{\cal F}_{\mu\nu}^2
-\frac{4i\Theta}{3}\hat L(\Sigma, C)\right]
\int {\cal D}x^{(a)}(\xi)
\delta\left(\sum\limits_{a=1}^{3}
\Sigma_{\mu\nu}^a\right)\times$$

\begin{equation}
\label{result}
\times {\cal D}h_{\mu\nu}^a\exp\Biggl\{\int d^4x\Biggl[-
\frac{1}{24\eta^2}\left(H_{\mu\nu\lambda}^a\right)^2-
\frac{3g_m^2}{8}\left(h_{\mu\nu}^a\right)^2+
i\pi h_{\mu\nu}^a\bar\Sigma_{\mu\nu}^a
\Biggr]\Biggr\}.
\end{equation}
Here, 
$H_{\mu\nu\lambda}^a=\partial_\mu h_{\nu\lambda}^a+
\partial_\lambda h_{\mu\nu}^a+\partial_\nu h_{\lambda\mu}^a$ 
is the field strength tensor of 
the Kalb-Ramond field of the $a$-th type, $h_{\mu\nu}^a$, 
${\cal F}_{\mu\nu}$ is defined by Eq.~(\ref{calF}), and  
$\hat L(\Sigma, C)\equiv\frac{1}{8\pi^2}\varepsilon_{\mu\nu\lambda\rho}
\int d^4xd^4y\Sigma_{\mu\nu}(x)j_\rho(y)\frac{\partial}{\partial x_\lambda}
\frac{1}{(x-y)^2}$
is the (formal expression for the) 4D Gauss linking number
of the surface $\Sigma$ with its boundary $C$, which will be shown to
become cancelled from the 
final expression for the partition function.
In Eq.~(\ref{result}), 
we have also denoted $\bar\Sigma_{\mu\nu}^a\equiv\Sigma_{\mu\nu}^a-
s_a^{(c)}\Sigma_{\mu\nu}-
\frac{i\Theta g_m^2}{4\pi^2}s_a^{(c)}\tilde{\cal F}_{\mu\nu}$
and 
introduced the following numbers 
$s_a^{(c)}$'s: $s_3^{(R)}=s_2^{(B)}=
s_1^{(G)}=0$, $s_1^{(R)}=s_3^{(B)}=s_2^{(G)}=
-s_2^{(R)}=-s_1^{(B)}=-s_3^{(G)}=1$, which obey the relation
${\bf Q}^{(c)}=\frac{g}{3}{\bf e}_as_a^{(c)}$.

As the next step, one needs to carry out the integration over the 
Kalb-Ramond fields $h_{\mu\nu}^a$'s. Referring the reader for 
details of this procedure to Appendix B, we shall present here 
only the outcome of the respective calculation, which has the form 

$$
\int {\cal D}h_{\mu\nu}^a\exp\Biggl\{\int d^4x\Biggl[-
\frac{1}{24\eta^2}\left(H_{\mu\nu\lambda}^a\right)^2-
\frac{3g_m^2}{8}\left(h_{\mu\nu}^a\right)^2+
i\pi h_{\mu\nu}^a\bar\Sigma_{\mu\nu}^a
\Biggr]\Biggr\}=$$

\begin{equation}
\label{KR}
=\exp\left\{-2\pi^2\int d^4xd^4y{\cal D}_m^{(4)}(x-y)\left[\eta^2
\bar\Sigma_{\mu\nu}^a(x)
\bar\Sigma_{\mu\nu}^a(y)
+\frac{4}{3g_m^2}j_\mu(x)j_\mu(y)\right]\right\}.
\end{equation}
Here, ${\cal D}_m^{(4)}(x)\equiv
\frac{m}{4\pi^2}\frac{K_1(m|x|)}{|x|}$ 
is the propagator of a dual vector 
boson of the mass $m=g_m\eta\sqrt{3}$ 
with $K_1$ standing 
for the modified Bessel function. 

Further, it is necessary to carry out the integral

\begin{equation}
\label{int1}
2\left(\frac{\Theta g_m^2\eta}{4\pi}\right)^2\left(s_a^{(c)}\right)^2
\int d^4x d^4y\tilde{\cal F}_{\mu\nu}(x){\cal D}_m^{(4)}(x-y)
\tilde{\cal F}_{\mu\nu}(y),
\end{equation}
emerging among other terms in the expression
$-2\pi^2\eta^2\int d^4xd^4y
\bar\Sigma_{\mu\nu}^a(x)
{\cal D}_m^{(4)}(x-y)\bar\Sigma_{\mu\nu}^a(y)$, which stands 
in the argument of the exponent on the R.H.S. of Eq.~(\ref{KR}).  
This calculation is outlined in Appendix C, and the result reads 

\begin{equation}
\label{int2}
\frac16\left(\frac{\Theta g_m}{2\pi^2}\right)^2\int d^4x d^4y
j_\mu(x)\left[\frac{1}{(x-y)^2}-4\pi^2{\cal D}_m^{(4)}(x-y)
\right]j_\mu(y).
\end{equation}
On the other hand, after substituting Eq.~(\ref{calF}) into
the term 
$-\frac{(\Theta g_m)^2}{12\pi^2}\int d^4x
{\cal F}_{\mu\nu}^2$, which stands in the first exponent on the R.H.S. of 
Eq.~(\ref{result}), we get for this term the expression 
$-\frac{(\Theta g_m)^2}{24\pi^4}\int d^4x d^4y j_\mu(x)
\frac{1}{(x-y)^2}j_\mu(y)$, which precisely cancels the corresponding 
Coulomb interaction of currents in Eq.~(\ref{int2}). Clearly, 
this cancellation could be anticipated from the very beginning just from 
physical principles, since in the resulting theory there does not 
remain any massless particles, by which the quarks might exchange.

Substituting now Eqs.~(\ref{KR}) and~(\ref{int2}) into Eq.~(\ref{result}),
we arrive at the following intermediate result for the partition 
function

$${\cal Z}_c=
\exp\left[
-\frac{4i\Theta}{3}\hat L(\Sigma, C)
-\left(\frac{8\pi^2}{3g_m^2}+\frac{(\Theta g_m)^2}{6\pi^2}\right)
\int d^4x d^4y j_\mu(x){\cal D}_m^{(4)}(x-y)j_\mu(y)\right]
\times$$

$$
\times\int {\cal D}x^{(a)}(\xi)
\delta\left(\sum\limits_{a=1}^{3}
\Sigma_{\mu\nu}^a\right)
\exp\Biggl[
-2(\pi\eta)^2\int d^4x d^4y\hat\Sigma_{\mu\nu}^a(x)
{\cal D}_m^{(4)}(x-y)\hat\Sigma_{\mu\nu}^a(y)+$$

\begin{equation}
\label{inter}
+i\Theta g_m^2\eta^2
s_a^{(c)}\int d^4x d^4y\hat\Sigma_{\mu\nu}^a(x){\cal D}_m^{(4)}(x-y)
\tilde{\cal F}_{\mu\nu}(y)\Biggr],
\end{equation}
where it has been denoted $\hat\Sigma_{\mu\nu}^a\equiv\Sigma_{\mu\nu}^a-
s_a^{(c)}\Sigma_{\mu\nu}$. Clearly, the last term in the argument of the 
second exponent on the R.H.S. of Eq.~(\ref{inter}) requires further 
simplifications. By virtue of Eq.~(\ref{calF}) and a partial integration, 
this term can be rewritten as 

$$i\Theta g_m^2\eta^2\frac{m}{(4\pi^2)^2}s_a^{(c)}
\varepsilon_{\mu\nu\lambda\rho}\int d^4xd^4yd^4z
\hat\Sigma_{\mu\nu}^a(x)\frac{j_\lambda(z)}{(y-z)^2}
\frac{\partial}{\partial x_\rho}\frac{K_1(m|x-y|)}{|x-y|}.$$
The integral over $y$ has the same form as the integral $I(\lambda)$ from 
Appendix C with $\lambda=x-z$ 
and reads $\frac{4\pi^2}{m^3}\left[\frac{1}{(x-z)^2}-
4\pi^2{\cal D}_m^{(4)}(x-z)\right]$. Owing to this result and the 
definition of $\hat\Sigma_{\mu\nu}^a$, the term under study
takes the form 

\begin{equation}
\label{term}
-\frac{i\Theta}{3}\left[2s_a^{(c)}\hat L(\Sigma^a, C)-
4\hat L(\Sigma, C)-s_a^{(c)}\varepsilon_{\mu\nu\lambda\rho}
\int d^4xd^4y{\cal D}_m^{(4)}(x-y)j_\mu(x)\frac{\partial}{\partial y_\nu}
\hat\Sigma_{\lambda\rho}^a(y)\right].
\end{equation}
One can now see that, as it was expected, the singular term 
$-\frac{4i\Theta}{3}\hat L(\Sigma, C)$ from the first exponent 
on the R.H.S. of Eq.~(\ref{inter}) becomes cancelled by 
the corresponding term from Eq.~(\ref{term}). Finally, we arrive at 
the following expression for the partition function

$$
{\cal Z}_c=
\exp\left[
-\left(\frac{8\pi^2}{3g_m^2}+\frac{(\Theta g_m)^2}{6\pi^2}\right)
\int d^4x d^4y j_\mu(x){\cal D}_m^{(4)}(x-y)j_\mu(y)\right]
\int {\cal D}x^{(a)}(\xi)
\delta\left(\sum\limits_{a=1}^{3}
\Sigma_{\mu\nu}^a\right)\times
$$

$$
\times\exp\Biggl[
-2(\pi\eta)^2\int d^4x d^4y\hat\Sigma_{\mu\nu}^a(x)
{\cal D}_m^{(4)}(x-y)\hat\Sigma_{\mu\nu}^a(y)+
$$

\begin{equation}
\label{main}
+\frac{i\Theta}{3}s_a^{(c)}
\varepsilon_{\mu\nu\lambda\rho}
\int d^4xd^4y{\cal D}_m^{(4)}(x-y)j_\mu(x)\frac{\partial}{\partial y_\nu}
\hat\Sigma_{\lambda\rho}^a(y)-
\frac{2i\Theta}{3}s_a^{(c)}\hat L(\Sigma^a, C)\Biggr],
\end{equation}
which is the main result of the present Letter. Notice that  
for every colour $c$, it is straightforward to 
integrate one of 
the world-sheets $\Sigma^a$'s out of Eq.~(\ref{main})  
by resolving the constraint 
imposed by the corresponding $\delta$-function. This procedure
leads to an obvious  
expression for the partition function 
in terms of two independent string world-sheets.

The first exponent 
on the R.H.S. of Eq.~(\ref{main}) represents a short-ranged interaction 
of quarks, which due to the $\Theta$-term 
have now acquired also a nonvanishing magnetic charge,
{\it i.e.} became dyons. 
Clearly, the first term in the 
second exponent on the R.H.S. of Eq.~(\ref{main}) is the 
short-ranged (self-)interaction of four different world-sheets: 
three closed ones $\Sigma^a$'s and an open one $\Sigma$.   
This term is responsible for the linearly rising 
potential, which confines dyons. Upon the expansion in 
powers of the derivatives {\it w.r.t.} $\xi^a$'s, this term 
yields the coupling constants of the local 
string effective action (see the first paper from Ref.~\cite{su3} 
for details).
The second term is just the short-range interaction 
of dyons and strings. The most nontrivial term in the obtained 
expression is the last, third one, which describes a long-range 
interaction of dyons with closed world-sheets. Such an interaction 
represents the 4D-analogue of the Aharonov-Bohm effect~\cite{four}, 
which means that at $\Theta\ne 3\pi\times{\,}{\rm integer}$
({\it cf.} the explicit form of the numbers $s_a^{(c)}$'s), the 
dyons become scattered by the closed electric ANO strings.
Contrary to that, these critical values of $\Theta$ 
correspond to such a relation 
between the magnetic charge of a dyon and the electric flux inside the 
string when the scattering is absent.

\section{Acknowledgments}

The author is indebted to Prof. A. Di Giacomo for useful discussions and 
cordial hospitality. 
He is also greatful to the colleagues of the Quantum Field Theory Division
of the University of Pisa for kind hospitality and INFN for  
financial support.

\section{Appendices}

\subsection*{A. Path-Integral Duality Transformation}

In the present Appendix, we shall outline some details of a derivation 
of Eq.~(\ref{result}) from the main text. To proceed with, 
we shall firstly rewrite a $\theta_a^{\rm reg.}$-dependent part 
of the statistical
weight, entering 
the partition function~(\ref{partf}), as follows 

$$
\int {\cal D}\theta_a^{\rm reg.}{\cal D}k\exp\left\{\int d^4x\left[
-\eta^2\sum\limits_{a=1}^{3}
\left(\partial_\mu\theta_a-g_m{\bf e}_a{\bf B}_\mu\right)^2+
ik\sum\limits_{a=1}^{3}\theta_a^{\rm reg.}\right]\right\}=
$$

$$
=\int {\cal D}\theta_a^{\rm reg.} {\cal D}k 
{\cal D}C_\mu^a\exp\Biggl\{\int d^4x
\Biggl[-\frac{1}{4\eta^2}\left(C_\mu^a\right)^2+iC_\mu^a\left(
\partial_\mu\theta_a-g_m{\bf e}_a{\bf B}_\mu\right)
+ik\sum\limits_{a=1}^{3}\theta_a^{\rm reg.}\Biggr]\Biggr\} 
$$
and carry out the integration over $\theta_a^{\rm reg.}$'s. 
In this way, one needs to solve the equation $\partial_\mu C_\mu^a=k$, 
which should hold for an arbitrary index $a$. 
The solution to this equation reads

$$
C_\mu^a(x)=\partial_\nu\tilde
h_{\mu\nu}^a(x)-\frac{1}{4\pi^2}\frac{\partial}{\partial x_\mu}
\int d^4y\frac{k(y)}{(x-y)^2},$$
where $h_{\mu\nu}^a$ stands for the Kalb-Ramond field of the 
$a$-th type. 

Secondly, 
replacing the integrals over $\theta_a^{\rm sing.}$'s 
with the integrals over $x^{(a)}(\xi)$'s by virtue of Eq.~(\ref{suz3})   
and discarding for simplicity the 
Jacobians~\cite{polikarp} emerging during such 
changes of the 
integration variables~\footnote{Alternatively, these Jacobians can 
be referred to the integration measures ${\cal D}x^{(a)}$'s.}, we arrive 
at the following representation for the $\theta_a$-dependent part of the 
partition function~(\ref{partf})

$$
\int 
{\cal D}\theta_a^{\rm sing.}
{\cal D}\theta_a^{\rm reg.} {\cal D}k\delta\left(\sum\limits_{a=1}^{3}
\theta_a^{\rm sing.}\right)
\exp\Biggl\{\int d^4x\Biggl[
-\eta^2\sum\limits_{a=1}^{3}
\left(\partial_\mu\theta_a-g_m{\bf e}_a{\bf B}_\mu\right)^2+
ik\sum\limits_{a=1}^{3}\theta_a^{\rm reg.}\Biggr]\Biggr\}=$$

$$=\int {\cal D}k\exp\Biggl\{\frac{1}{4\pi^2}\int d^4x d^4y\Biggl[
-\frac{3}{4\eta^2}
\frac{k(x)k(y)}{(x-y)^2}+ig_m
\left(\frac{\partial}{\partial x_\mu}
\frac{k(y)}{(x-y)^2}\right)\sum\limits_{a=1}^{3}
{\bf e}_a{\bf B}_\mu(x) 
\Biggr]\Biggr\}\times$$

$$
\times\int {\cal D}x^{(a)}(\xi)\delta\left(\sum\limits_{a=1}^{3}
\Sigma_{\mu\nu}^a\right) {\cal D}h_{\mu\nu}^a
\exp\Biggl\{\int d^4x\Biggl[-\frac{1}{24\eta^2}
\left(H_{\mu\nu\lambda}^a\right)^2+i\pi h_{\mu\nu}^a\Sigma_{\mu\nu}^a-
ig_m{\bf e}_a
{\bf B}_\mu\partial_\nu \tilde h_{\mu\nu}^a\Biggr]\Biggr\}.
\eqno(A.1)
$$
Here, $H_{\mu\nu\lambda}^a=\partial_\mu h_{\nu\lambda}^a+
\partial_\lambda h_{\mu\nu}^a+\partial_\nu h_{\lambda\mu}^a$ stands 
for the field strength tensor of the Kalb-Ramond field $h_{\mu\nu}^a$. 
Clearly, due to the explicit form of the root vectors, 
the sum $\sum\limits_{a=1}^{3}{\bf e}_a{\bf B}_\mu$ vanishes, and 
the integration over the Lagrange multiplier $k$ thus yields an 
inessential (field-independent) determinant factor. Notice also that due to 
Eq.~(\ref{suz3}),  
the constraint 
$\sum\limits_{a=1}^{3}\theta_a^{\rm sing.}=0$ has gone over into a  
constraint for the world-sheets of closed electric ANO strings of three types
$\sum\limits_{a=1}^{3}\Sigma_{\mu\nu}^a=0$. This means that actually 
only the world-sheets of two types are independent of each other, 
whereas the third one is unambiguously fixed by the demand that the above 
constraint holds. 

Let us now turn ourselves to the pure gauge field sector of the 
partition function~(\ref{partf}). In this way, firstly the term 
$-\frac12\int d^4x{\bf F}_{\mu\nu}{\bf F}_{\mu\nu}^{(c)}$ can 
be rewritten as ${\bf Q}^{(c)}\int d^4x{\bf B}_\mu\partial_\nu
\tilde\Sigma_{\mu\nu}$. Secondly, the $\Theta$-term is equal 
to $\frac{i\Theta g_m^2}{8\pi^2}\varepsilon_{\mu\nu\lambda\rho}
\int d^4x{\bf B}_\mu\partial_\nu{\bf F}_{\lambda\rho}^{(c)}$.
(It is worth noting that 
this expression can further be written as 
$-\frac{i\Theta g_m}{\pi}
{\bf q}^{(c)}\int d^4x {\bf B}_\mu j_\mu$ 
with ${\bf q}^{(c)}\equiv
\frac{{\bf Q}^{(c)}}{g}$ being just a $g$-independent vector. 
The last equation means that due to the $\Theta$-term 
the quarks acquire a nonvanishing magnetic charge, {\it i.e.} become 
dyons. The 
$\Theta$-term then describes
an interaction of a dyon with the dual gauge field~\cite{witten}.)
After that, the gauge field sector takes the form

$$\int {\cal D}{\bf B}_\mu
\exp\Biggl\{\int d^4x\Biggl[
-\frac14\left({\bf F}_{\mu\nu}+{\bf F}_{\mu\nu}^{(c)}\right)^2
+\frac{i\Theta g_m^2}{16\pi^2}
\left({\bf F}_{\mu\nu}+{\bf F}_{\mu\nu}^{(c)}\right)
\left(\tilde{\bf F}_{\mu\nu}+\tilde{\bf F}_{\mu\nu}^{(c)}\right)
-ig_m{\bf e}_a
{\bf B}_\mu\partial_\nu \tilde h_{\mu\nu}^a
\Biggr]\Biggr\}=
$$

$$
=\int {\cal D}{\bf B}_\mu\exp\Biggl\{-\int d^4x\Biggl[
\frac14{\bf F}_{\mu\nu}^2+\frac14\left({\bf F}_{\mu\nu}^{(c)}\right)^2+
\varepsilon_{\mu\nu\lambda\rho}{\bf B}_\mu\partial_\nu\left(
\frac{ig_m}{2}{\bf e}_ah_{\lambda\rho}^a-
\frac{{\bf Q}^{(c)}}{2}\Sigma_{\lambda\rho}-
\frac{i\Theta g_m^2}{8\pi^2}{\bf F}_{\lambda\rho}^{(c)}\right)\Biggr]
\Biggr\}.\eqno(A.2)$$
It is further convenient to pass from the integration over the 
${\bf B}_\mu$-fields to the integration over the fields $B_\mu^a\equiv
{\bf e}_a{\bf B}_\mu$. Then, performing the rescaling $B_\mu^{a{\,}
{\rm (new)}}=
\sqrt{\frac23}B_\mu^{a{\,}{\rm (old)}}$ (in order to restore the standard 
factor $1/4$ in front of $\left(F_{\mu\nu}^a\right)^2$) and taking into 
account that 
${\bf Q}^{(c)}=\frac{g}{3}{\bf e}_as_a^{(c)}$ (see the notations 
after Eq.~(\ref{result})), we get for Eq.~(A.2) the following expression 

$$\int {\cal D}B_\mu^a\exp\Biggl\{-\int d^4x\Biggl[
\frac14\left(F_{\mu\nu}^a\right)^2+\frac14\left({\bf F}_{\mu\nu}^{(c)}
\right)^2+$$

$$+\frac12\sqrt{\frac32}\varepsilon_{\mu\nu\lambda\rho}B_\mu^a\partial_\nu
\left(ig_mh_{\lambda\rho}^a-\frac{g}{3}s_a^{(c)}\Sigma_{\lambda\rho}
-\frac{i\Theta g_m}{3\pi}s_a^{(c)}\tilde{\cal F}_{\lambda\rho}
\right)\Biggr]\Biggr\}.$$
After that, the integration over the $B_\mu^a$-fields can be carried out 
by making use of the first-order formalism, {\it i.e.} by representing
the factor $\exp\left[-\frac14\int d^4x\left(F_{\mu\nu}^a\right)^2\right]$
as an integral over an auxiliary antisymmetric tensor field as follows 

$$
\exp\left[-\frac14\int d^4x\left(F_{\mu\nu}^a\right)^2\right]=
\int {\cal D}G_{\mu\nu}^a
\exp\left\{\int d^4x\left[-\left(G_{\mu\nu}^a\right)^2+
i\varepsilon_{\mu\nu\lambda\rho}B_\mu^a\partial_\nu G_{\lambda\rho}^a
\right]\right\}.\eqno(A.3)
$$
The $B_\mu^a$-integration then yields the following expression for the 
field $G_{\mu\nu}^a$

$$G_{\mu\nu}^a=\frac12\sqrt{\frac32}\left[g_mh_{\mu\nu}^a+
\frac{ig}{3}s_a^{(c)}\Sigma_{\mu\nu}-\frac{\Theta g_m}{3\pi}s_a^{(c)}
\tilde{\cal F}_{\mu\nu}\right]+\partial_\mu
\lambda^a_\nu-\partial_\nu\lambda^a_\mu$$
with $\lambda^a_\mu$'s standing for the new fields, resulting from the 
resolution of the corresponding constraint.

Bringing now together Eqs.~(A.1) and~(A.3), we arrive at 
the following expression for the full partition function

$${\cal Z}_c=
\int {\cal D}x^{(a)}(\xi)\delta\left(\sum\limits_{a=1}^{3}
\Sigma_{\mu\nu}^a\right) {\cal D}h_{\mu\nu}^a{\cal D}\lambda^a_\mu 
\exp\Biggl\{\int d^4x\Biggl[-\frac{1}{24\eta^2}
\left(H_{\mu\nu\lambda}^a\right)^2-$$

$$
-\left[\frac12\sqrt{\frac32}\left(g_mh_{\mu\nu}^a+
\frac{ig}{3}s_a^{(c)}\Sigma_{\mu\nu}-\frac{\Theta g_m}{3\pi}s_a^{(c)}
\tilde{\cal F}_{\mu\nu}\right)+\partial_\mu
\lambda^a_\nu-\partial_\nu\lambda^a_\mu\right]^2
+i\pi h_{\mu\nu}^a\Sigma_{\mu\nu}^a-\frac14\left({\bf F}_{\mu\nu}^{(c)}
\right)^2\Biggr]\Biggr\}.\eqno(A.4)$$
Clearly, due to the closeness of the world-sheets $\Sigma^a$'s, the obtained 
action standing in the exponent on the R.H.S. of Eq.~(A.4) is 
hypergauge-invariant, {\it i.e.} it is invariant under the transformations
$h_{\mu\nu}^a\to h_{\mu\nu}^a+\partial_\mu\gamma^a_\nu-
\partial_\nu\gamma^a_\mu$, $\lambda^a_\mu\to\lambda^a_\mu-\frac{g_m}{2}
\sqrt{\frac32}\gamma^a_\mu$. Owing to that, the fields $\lambda^a_\mu$'s 
can be completely eliminated by setting $\gamma^a_\mu=\frac{2}{g_m}
\sqrt{\frac23}\lambda^a_\mu$, which leads to the following expression 

$$
{\cal Z}_c=\int {\cal D}x^{(a)}(\xi)\delta\left(\sum\limits_{a=1}^{3}
\Sigma_{\mu\nu}^a\right){\cal D}h_{\mu\nu}^a\exp\Biggl\{-\int d^4x\Biggl[
\frac{1}{24\eta^2}\left(H_{\mu\nu\lambda}^a\right)^2+$$

$$
+\frac38\left(
g_mh_{\mu\nu}^a+\frac{ig}{3} s_a^{(c)}
\Sigma_{\mu\nu}-\frac{\Theta g_m}{3\pi} s_a^{(c)}
\tilde{\cal F}_{\mu\nu}\right)^2
-i\pi h_{\mu\nu}^a\Sigma_{\mu\nu}^a+\frac14\left({\bf F}_{\mu\nu}^{(c)}
\right)^2\Biggr]\Biggr\}.\eqno(A.5)
$$
Next, a straightforward algebra yields 

$$
\frac38\left(
g_mh_{\mu\nu}^a+\frac{ig}{3} s_a^{(c)}
\Sigma_{\mu\nu}-\frac{\Theta g_m}{3\pi} s_a^{(c)}
\tilde{\cal F}_{\mu\nu}\right)^2=$$

$$=\frac{(\Theta g_m)^2}{12\pi^2}{\cal F}_{\mu\nu}^2+
\frac{3g_m^2}{8}\left(h_{\mu\nu}^a\right)^2-\frac{g^2}{12}
\Sigma_{\mu\nu}^2+i\pi s_a^{(c)}h_{\mu\nu}^a\Sigma_{\mu\nu}-
\frac{\Theta g_m^2}{4\pi}s_a^{(c)}h_{\mu\nu}^a\tilde{\cal F}_{\mu\nu}-
\frac{2i\Theta}{3}\Sigma_{\mu\nu}\tilde{\cal F}_{\mu\nu},$$
where we have taken into account that for 
every $c$, $\left(s_a^{(c)}\right)^2=2$.
Clearly, the singular term
$-\frac14\left({\bf F}_{\mu\nu}^{(c)}
\right)^2$ becomes cancelled by the term $\frac{g^2}{12}
\Sigma_{\mu\nu}^2$. Finally, substituting for ${\cal F}_{\mu\nu}$ 
Eq.~(\ref{calF}), we can rewrite 
the term 
$\frac{2i\Theta}{3}\int d^4x\Sigma_{\mu\nu}\tilde{\cal F}_{\mu\nu}$ as
$-\frac{4i\Theta}{3}\hat L(\Sigma, C)$, where 
$\hat L(\Sigma, C)\equiv\frac{1}{8\pi^2}\varepsilon_{\mu\nu\lambda\rho}
\int d^4xd^4y\Sigma_{\mu\nu}(x)j_\rho(y)\frac{\partial}{\partial x_\lambda}
\frac{1}{(x-y)^2}$
is the formal expression for the 
4D Gauss linking number of the surface $\Sigma$ with its boundary
$C$. As it will be demonstrated below, this singular 
term eventually cancels out.

Accounting for the above considerations in Eq.~(A.5), we arrive at 
Eq.~(\ref{result}) of the main text.

\subsection*{B. Integration over the Kalb-Ramond Fields}

In this Appendix, we shall present some details of a derivation 
of Eq.~(\ref{KR}) from the main text. Namely, we shall carry out the 
following functional integral over the Kalb-Ramond fields

$$
\int {\cal D}h_{\mu\nu}^a\exp\Biggl[
-\int d^4x\Biggl(\frac{1}
{24\eta^2}
\left(H_{\mu\nu\lambda}^a\right)^2+\frac{3g_m^2}{8}
\left(h_{\mu\nu}^a\right)^2-
i\pi h_{\mu\nu}^a\bar\Sigma_{\mu\nu}^a
\Biggr)\Biggr].\eqno(B.1) 
$$
Clearly, such a Gaussian integral can be calculated upon 
the substitution of its saddle-point value back into the action.
The saddle-point equation in the momentum 
representation reads 

$$\frac{1}{4\eta^2}\left(p^2h_{\nu\lambda}^{a{\,}{\rm (extr.)}}(p)+p_\lambda 
p_\mu h_{\mu\nu}^{a{\,}{\rm (extr.)}}(p)+
p_\mu p_\nu h_{\lambda\mu}^{a{\,}{\rm (extr.)}}(p)
\right)+\frac{3g_m^2}{4}h_{\nu\lambda}^{a{\,}{\rm (extr.)}}(p)=i\pi
\bar\Sigma_{\nu\lambda}^a(p).$$
This equation can most easily be solved by rewriting it in the 
following way

$$
\left(p^2 \hat P_{\lambda\nu, \alpha\beta}+m^2 \hat 1_{\lambda\nu, 
\alpha\beta}\right)h_{\alpha\beta}^{a{\,}{\rm (extr.)}}(p)=4\pi i\eta^2
\bar\Sigma_{\lambda
\nu}^a(p),\eqno(B.2) 
$$
where $m=g_m\eta\sqrt{3}$ is just the mass of the dual gauge bosons,
equal to the mass of the Kalb-Ramond fields.
In Eq.~(B.2), we have also introduced the following projection operators

$$\hat P_{\mu\nu, \lambda\rho}\equiv\frac12\left({\cal P}_{\mu\lambda}
{\cal P}_{\nu\rho}-{\cal P}_{\mu\rho} {\cal P}_{\nu\lambda}\right)~
\mbox{and}~ 
\hat 1_{\mu\nu, \lambda\rho}\equiv\frac12 \left(\delta_{\mu\lambda}
\delta_{\nu\rho}-\delta_{\mu\rho}\delta_{\nu\lambda}\right)$$
with ${\cal P}_{\mu\nu}\equiv\delta_{\mu\nu}-\frac{p_\mu p_\nu}{p^2}$. 
These projection operators obey the following relations 

$$
\hat 1_{\mu\nu, \lambda\rho}=-\hat 1_{\nu\mu, \lambda\rho}=
-\hat 1_{\mu\nu, \rho\lambda}=\hat 1_{\lambda\rho, \mu\nu},~ 
\hat 1_{\mu\nu, \lambda\rho} \hat 1_{\lambda\rho, \alpha\beta}=
\hat 1_{\mu\nu, \alpha\beta}\eqno(B.3) 
$$
(the same relations hold for $\hat P_{\mu\nu, \lambda\rho}$), and 

$$
\hat P_{\mu\nu, \lambda\rho}\left(\hat 1-\hat P\right)_{\lambda
\rho, \alpha\beta}=0.\eqno(B.4) 
$$
By virtue of the properties~(B.3) and~(B.4), 
the solution to the saddle-point equation~(B.2) reads

$$h_{\lambda\nu}^{a{\,}{\rm (extr.)}}(p)=\frac{4\pi i\eta^2}{p^2+m^2}\left[
\hat 1+\frac{p^2}{m^2}\left(\hat 1-\hat P\right)\right]_{\lambda
\nu, \alpha\beta}\bar\Sigma_{\alpha\beta}^a(p),$$
which, once being substituted back into the functional integral~(B.1), 
yields for it the following expression

$$
\exp\Biggl\{-2\pi^2\eta^2\int\frac{d^4p}{(2\pi)^4}
\frac{1}{p^2+m^2}\left[\hat 1+\frac{p^2}{m^2}\left(\hat 1-\hat P
\right)\right]_{\mu\nu, \alpha\beta}\bar\Sigma_{\mu\nu}^a(-p)
\bar\Sigma_{\alpha\beta}^a(p)
\Biggr\}.\eqno(B.5) 
$$
Rewriting now the term from  
Eq.~(B.5), which is 
proportional to the 
projection operator $\hat 1$,
in the coordinate representation we obtain 
the expression 
$2\pi^2\eta^2\int d^4x d^4y\bar\Sigma_{\mu\nu}^a(x)
{\cal D}_m^{(4)}(x-y)
\bar\Sigma_{\mu\nu}^a(y)$ 
from the action standing in the exponent 
on the R.H.S. of Eq.~(\ref{KR}).

Let us now consider the term 
proportional to the projection operator 
$\left(\hat 1-\hat P\right)$ on the 
R.H.S. of Eq.~(B.5). In this way, by making use of the following equation

$$
p^2(\hat 1-\hat P)_{\mu\nu, \alpha\beta}=\frac12
(\delta_{\nu\beta}p_\mu p_\alpha+\delta_{\mu\alpha}
p_\nu p_\beta-\delta_{\nu\alpha}p_\mu p_\beta-\delta_{\mu
\beta}p_\nu p_\alpha),
$$
we see that the term under study, 

$$-2\pi^2\eta^2\int\frac{d^4p}{(2\pi)^4}\frac{1}{p^2+m^2}
\frac{p^2}{m^2}\left(\hat 1-\hat P\right)_{\mu\nu, \alpha\beta}
\int d^4x d^4y
{\rm e}^{ip(y-x)}\bar\Sigma_{\mu\nu}^a(x)\bar\Sigma_{\alpha\beta}^a(y),$$
after carrying out the integration over $p$, reads

$$
\left(\frac{2\pi\eta}{m}\right)^2\int d^4x d^4y\bar\Sigma_{\mu\nu}^a(x)
\bar\Sigma_{\nu\beta}^a(y) 
\frac{\partial^2}{\partial x_\mu\partial y_\beta}
{\cal D}_m^{(4)}(x-y).
$$
Performing twice the partial integration and making use of the 
equation $\partial_\mu\bar\Sigma_{\mu\nu}^a=s_a^{(c)}j_\nu$, 
we can further rewrite this expression as 
$-\frac{8\pi^2}{3g_m^2}
\int d^4x d^4y j_\mu(x)
{\cal D}_m^{(4)}(x-y)j_\mu(y)$, which is just the last term 
in the exponent on the R.H.S. of Eq.~(\ref{KR}).

\subsection*{C. Calculation of the Integral~(\ref{int1})}

Below in this Appendix, we shall present some details of a calculation 
of the integral~(\ref{int1}). 
Owing to Eq.~(\ref{calF}) 
and the fact that for every $c$, $\left(s_a^{(c)}\right)^2=2$, 
this integral can be 
rewritten as follows 

$$\frac{m}{8\pi^4}\left(\frac{\Theta g_m^2\eta}{4\pi^2}\right)^2
\int d^4x d^4y\frac{K_1(m|x-y|)}{|x-y|}\frac{\partial}{\partial x_\mu}
\int d^4z\frac{j_\nu(z)}{(x-z)^2}\frac{\partial}{\partial y_\mu}
\int d^4u\frac{j_\nu(u)}{(y-u)^2}.\eqno(C.1)$$
Next, by making use of the following sequence of partial integrations

$$ 
\frac{\partial}{\partial y_\mu}\frac{1}{(y-u)^2}\to
-\frac{\partial}{\partial y_\mu}\frac{K_1(m|x-y|)}{|x-y|}=
\frac{\partial}{\partial x_\mu}\frac{K_1(m|x-y|)}{|x-y|}\to
-\frac{\partial}{\partial x_\mu}\frac{1}{(x-z)^2},$$
one can carry out the integration over $z$. After that, the 
integral~(C.1) takes the form

$$\frac{m}{2\pi^2}
\left(\frac{\Theta g_m^2\eta}{4\pi^2}\right)^2\int d^4x d^4u j_\mu(x)
I(x-u) j_\mu(u),\eqno(C.2)$$
where $I(x-u)\equiv\int d^4y\frac{K_1(m|x-y|)}{|x-y|(y-u)^2}$ is 
just the integral, which is left to be calculated.

In order to proceed with this calculation, let us pass to the 
integration over the variable $z\equiv y-u$ 
and introduce the vector $\lambda\equiv
x-u$, after which we obtain 
$I(\lambda)=\int d^4z\frac{K_1(m|z-\lambda|)}{|z-\lambda|}\frac{1}{z^2}$.
Such an integral can most easily be carried out by substituting for 
the two ratios, which form the integrand, the R.H.S.'s of the 
following equalities

$$\frac{K_1(m|z-\lambda|)}{|z-\lambda|}=\frac{4\pi^2}{m}
\int\frac{d^4p}{(2\pi)^4}\frac{{\rm e}^{ip(z-\lambda)}}{p^2+m^2}~
{\rm and}~ \frac{1}{z^2}=4\pi^2\int\frac{d^4q}{(2\pi)^4}
\frac{{\rm e}^{iqz}}{q^2}.$$
Then we have 

$$
I(\lambda)=
\frac{16\pi^4}{m}\int\frac{d^4p}{(2\pi)^4}
\frac{{\rm e}^{-ip\lambda}}{(p^2+m^2)p^2}=$$

$$=\frac{16\pi^4}{m}
\int\frac{d^4p}{(2\pi)^4}\int\limits_{0}^{+\infty}d\alpha
\int\limits_{0}^{+\infty}d\beta{\rm e}^{-ip\lambda-\alpha p^2-
\beta(p^2+m^2)}=\frac{\pi^2}{m}\int\limits_{0}^{+\infty}d\alpha
\int\limits_{0}^{+\infty}d\beta
\frac{{\rm e}^{-\beta m^2-
\frac{\lambda^2}{4(\alpha+\beta)}}}{(\alpha+\beta)^2}.\eqno (C.3)$$  
It is further convenient to introduce new integration variables 
$a\in [0,+\infty)$ and $t\in [0,1]$ according to the formulae 
$\alpha=at$ and $\beta=a(1-t)$. Straightforward integration over $t$ 
then reduces the integral~(C.3) to the following expression 

$$
I(\lambda)=\frac{\pi^2}{m^3}\int\limits_{0}^{+\infty}\frac{da}{a^2}
{\rm e}^{-\frac{\lambda^2}{4a}}\left(1-{\rm e}^{-am^2}\right).$$
Such an integral can be carried out by virtue  
of the formula

$$\int\limits_{0}^{+\infty}x^{\nu-1}
{\rm e}^{-\frac{\beta}{x}-\gamma x}dx=
2\left(\frac{\beta}{\gamma}\right)^{\frac{\nu}{2}}K_\nu\left(2
\sqrt{\beta\gamma}\right),~ \Re\beta>0,~ \Re\gamma>0,$$
and the result has the form 

$$
I(\lambda)=
\frac{4\pi^2}{m^3|\lambda|}\left[\frac{1}{|\lambda|}-
mK_1(m|\lambda|)\right].$$
Finally, substituting this expression into Eq.~(C.2), we arrive at 
Eq.~(\ref{int2}) of the main text.


\newpage


\begin{thebibliography}{99}

\bibitem{th}
G. 't Hooft, Nucl. Phys. {\bf B 190} (1981) 455.

\bibitem{digiacomo}
A. Di Giacomo, preprint {\tt hep-lat/9907010} (1999), 
preprint {\tt hep-lat/9907029} (1999).

\bibitem{abdom}
Z.F. Ezawa and A. Iwazaki, Phys. Rev. {\bf D 25} (1982) 2681; 
ibid. {\bf D 26} (1982) 631.

\bibitem{ano}
A.A. Abrikosov, Sov. Phys.- JETP {\bf 5} (1957) 1174;
H.B. Nielsen and P. Olesen, Nucl. Phys. {\bf B 61} (1973) 45;
for a review see {\it e.g.} E.M. Lifshitz and L.P. Pitaevski, 
{\it Statistical Physics, Vol. 2} (Pergamon, New York, 1987).

\bibitem{mand}
S. Mandelstam, Phys. Lett. {\bf B 53} (1975) 476; Phys. Rep. 
{\bf C 23} (1976) 245; G. 't Hooft, in: {\it High Energy Physics}, 
Ed. A. Zichichi (Editrice Compositori, Bologna, 1976).

\bibitem{maedan}
S. Maedan and T. Suzuki, Prog. Theor. Phys. {\bf 81} (1989) 229.

\bibitem{su3}
D. Antonov and D. Ebert, Phys. Lett. {\bf B 444} (1998) 208; 
D.A. Komarov and M.N. Chernodub, JETP Lett. {\bf 68} (1998) 117.

\bibitem{rev}
D. Antonov and D. Ebert, in: {\it Path Integrals from peV to TeV: 
50 years after Feynman's paper}, Eds. R. Casalbuoni {\it et al.} 
(World Scientific, Singapore, 1999), pp. 267-271; 
preprint {\tt CERN-TH/99-294}, {\tt hep-th/9909156} (1999) 
(Nucl. Phys. {\bf B} (Proc. Suppl.), in press).

\bibitem{mpla}
D. Antonov, Mod. Phys. Lett. {\bf A 14} (1999) 1829.

\bibitem{ijmpa}
D. Antonov, Int. J. Mod. Phys. {\bf A 14} (1999) 4347.

\bibitem{mono}
R. Rajaraman, {\it An Introduction to Solitons and Instantons 
in Quantum Field Theory} (North-Holland Publishing Company, 
Amsterdam, 1982); J.F. Donoghue, E. Golowich, and B.R. Holstein, 
{\it Dynamics of the Standard Model} (Cambridge Univ. Press, 
Cambridge, 1992); J. Zinn-Justin, {\it Quantum Field Theory and 
Critical Phenomena, 2nd ed.} (Oxford Univ. Press Inc., New York, 
1993).

\bibitem{witten}
E. Witten, Phys. Lett. {\bf B 86} (1979) 283.

\bibitem{four}
M.G. Alford, J. March-Russel, and F. Wilczek, Nucl. Phys. 
{\bf B 337} (1990) 695; J. Preskill and L.M. Krauss, Nucl. Phys. 
{\bf B 341} (1990) 50.


\bibitem{ab}
Y. Aharonov and D. Bohm, Phys. Rev. {\bf 115} (1959) 485.

\bibitem{emil}
E.T. Akhmedov, JETP Lett. {\bf 64} (1996) 82.

\bibitem{polikarp}
E.T. Akhmedov, M.N. Chernodub, M.I. Polikarpov, and M.A. Zubkov, 
Phys. Rev. {\bf D 53} (1996) 2087.

\bibitem{lee}
K. Lee, Phys. Rev. {\bf D 48} (1993) 2493;
P. Orland, Nucl. Phys. {\bf B 428} (1994) 221;
D. Antonov and D. Ebert, Eur. Phys. J. {\bf C 8} (1999) 343.


\end{thebibliography}
\end{document}